# Magnetic properties of dense nanoparticle arrays with core/shell morphology

D. Kechrakos* , K. N. Trohidou and M. Vasilakaki

*Institute of Materials Scienc, NCSR Demokritos, 153 10 Athens, Greece*

**Abstract**

We calculate the magnetization hysteresis for an ordered array of composite magnetic nanoparticles with a ferromagnetic (FM) core and an antiferromagnetic (AFM) shell, located on a triangular lattice and coupled via magnetostatic forces ($g$). Each nanoparticle is described by a pair of exchange-coupled ($J$), anisotropic spins ($K_C$, $K_S$) (Meiklejohn-Bean model). The magnetization hysteresis loop is obtained using the Metropolis Monte Carlo algorithm. For magnetically hard nanoparticles ($K_C \geq g$) we find that the coercivity is reduced with increasing the dipolar coupling strength, while the exchange bias field shows an non-monotonous behavior resulting from the competition between the random anisotropy and interparticle dipolar interactions. The possibility of enhancing the exchange bias field by increasing the packing density is discussed.

*PACS:* 75.50.Tt; 75.50.Ss; 75.75.+a

*Keywords:* magnetic nanoparticles; ordered arrays; exchange bias; dipolar interactions; Monte Carlo

## 1. Introduction

Cobalt nanoparticles surrounded by an oxide shell are the first systems discovered to exhibit the exchange bias effect [1]. More recently, the exchange bias effect in laterally confined structures (dots and nanoparticles) has attracted renewed interest [2] as exploitation of this effect holds promises for development of smaller magnetic structures with higher blocking temperature [3]. Studies of the exchange bias effect in core-shell nanostructures have aimed to relate macroscopic characteristics, such as the coercive field ($H_C$) and the exchange bias field ($H_E$), to the material parameters (anisotropy energy, exchange coupling strength), the structural parameters (interface microstructure, defects) and the underlying microscopic mechanism of magnetization reversal (domain wall formation). [2]. Despite the research effort focused on the microscopic details of the exchange bias mechanism in individual nanoparticles [4], much less attention has been paid so far to the modification of the magnetic hysteresis behavior due to inter-particle interactions arising in assemblies. In this direction, iron nanoparticles embedded in iron-oxide matrix [5] were shown to freeze below a temperature owing to the competition between the exchange anisotropy at the core-shell interface and the interparticle dipolar interactions. Similarly, increase of the exchange bias field due to magnetostatic interparticle coupling was found in stripes of Co/CoO nanoparticles [6]. Ordered arrays of magnetic nanoparticles prepared by chemical synthesis are currently promising materials for applications in high-density (~Tb/in$^2$) magnetic storage media, due to the sharp distribution of their magnetic properties and their high reproducibility [7, 8]. Synthesis of Fe-based or Co-based nanoparticles leads to formation of closed packed arrays of strongly dipolar interacting nanoparticles with core-shell morphology [8]. In previous studies [9, 10] we have demonstrated that the remanence and coercivity of dipolar coupled FM nanoparticle arrays vary strongly with surface coverage, interparticle distance and number of deposited monolayers. The modification of the coercive and exchange-bias fields in dense nanoparticle arrays with core-shell morphology as a result of the competition between exchange anisotropy and interparticle dipolar interactions consists a challenging question, in view of the exploitation of these arrays as hard magnetic materials. The aim of the present work is to address this problem implementing the simplified mesoscopic model of Meiklejohn and Bean (MB model) [1] to describe the magnetic structure of the core-shell nanoparticles. This model assumes coherent rotation of the core and shell magnetization, which is a reasonable first approximation for nanoparticles with a single-domain ground state (D≈10nm). It is well established [2] that the MB model produces unphysically large $H_E$ values when bulk values for the interface exchange coupling are used. Also, experiments [2] and atomic scale calculations [4] have provided ample evidence that the structural and magnetic properties at the core-shell interface play a crucial role in the exchange bias effect of nanostructures. We assume that to a first approximation, these microscopic details would modify the effective interface exchange coupling of the MB model and therefore, in the present study we let this coupling take different values.

---

* Corresponding author. Tel.: +310-210-650-3313; fax: +310-210-651-9430.
*E-mail address*: dkehrakos@ims.demokritos.gr.

**2. Model and Simulation Method**

We consider identical spherical particles with core diameter D and shell diameter $D_0$ forming a two-dimensional triangular lattice in the xy-plane with lattice constant $d \geq D_0$. In each nanoparticle, the core and shell have uniaxial anisotropy along a common easy axis, that points in a random direction, and anisotropy energies $K_C$ and $K_S$, respectively. The exchange coupling across the core-shell interface of the nanoparticle is described by an effective coupling parameter J (J>0). An externally applied magnetic field couples only to the FM spins as the total moment of the AFM shell is approximately zero [1]. Long-range dipolar interactions between the cores are included with strength $g=\mu^2/d^3$, where $\mu=M_s V_C$ is the magnetic moment of the core. Under these assumptions the total energy of the assembly is given as

$$E = -K_C \sum_i \left(\hat{S}_i^f \cdot \hat{e}_i\right)^2 - K_S \sum_i \left(\hat{S}_i^a \cdot \hat{e}_i\right)^2 - J \sum_i \left(\hat{S}_i^f \cdot \hat{S}_i^a\right)$$

$$-h \sum_i \left(\hat{S}_i^f \cdot \hat{H}\right) - g \sum_i \frac{\left(\hat{S}_i^f \cdot \hat{S}_j^f\right) - 3\left(\hat{S}_i^f \cdot \hat{R}_{ij}\right)\left(\hat{S}_j^f \cdot \hat{R}_{ij}\right)}{\left(R_{ij}/d\right)^3}$$

where superscripts "f" and "a" indicate the FM and AFM spins, respectively, and hats indicate unit vectors. We scale all energy parameters entering our simulations by the core anisotropy energy ($K_C$=1). The anisotropy of the AFM oxide is assumed much higher than the core anisotropy, and in the present work we take $K_S$=5.0. As discussed below, this value of $K_S$ is high enough to ensure the blocking of the shell magnetization ($S^a$) for applied fields in the range that the core magnetization ($S^f$) exhibits hysteresis behavior. For transition metal nanoparticles typical values of the anisotropy are $K_C \approx 10^6$ erg/cm$^3$ and the magnetization $M_s \approx 10^3$ emu/cm$^3$, which give for the dipolar coupling strength $g/K_C \approx 0.52(D/d)^3$. To avoid demagnetizing effects due to free boundaries we use periodic boundaries in the xy-plane and free boundaries in the z-axis. The dipolar sums are calculated using the Ewald method [11]. The cooling saturation field, required to observe the exchange bias effect is taken along the positive x-axis. Low-temperature (T=0.01 $K_C/k_B$) isothermal hysteresis loops are obtained with the applied field (H) along a symmetry axis of the triangular lattice (Ox) and values in the range $-3 \leq \mu H/K_C \leq +3$ using a step dH=0.1($K_C/\mu$). At each intermediate field value, thermalization of the system is allowed for the first $10^3$ Monte Carlo steps per spin (MCS) and thermal averages are calculated over the subsequent $10^4$ MCS, by sampling every 10 MCS in order to obtain uncorrelated data. Simulations are performed on a 20 x 20 cell and the results are averaged over 10 samples with different realizations of the easy axes distribution. In all loops presented below the magnetization is reduced to the total core volume. The exchange field ($H_E$) and the effective coercive field ($H_C$) are given as $H_E=|H_{C2}+H_{C1}|/2$ and $H_C=|H_{C2}-H_{C1}|/2$, where $H_{C1}$ and $H_{C2}$ are the left and right coercive fields, respectively.

**3. Results and Discussion**

In fig.1 we plot the isothermal hysteresis loop of a two-dimensional nanoparticle assembly for various dipolar coupling strengths, or equivalently, for various interparticle distances (g ~ $1/d^3$) ranging from contact (g=0.5) to infinite separation (g=0). For comparison reasons, we show data for two different nanoparticle assemblies with moderate (J=1.0) and strong (J=1.5) interface exchange coupling. We discuss the changes in the magnetization curves introduced by the presence of interparticle interactions. First, the remanence increases with increasing dipolar strength owing to the ferromagnetic character of dipolar interactions in the triangular lattice. Second, an overall change of the loop shape to a square-like one is seen. This feature indicates a sudden switch of the magnetization, which is characteristic of a coherent rotation of the magnetization under an applied field along the easy axis of the system. In the case of dipoles lying on triangular lattice dipolar interactions introduce a uniaxial anisotropy along the symmetry axes of the lattice, that leads to the observed change of loop shape. Third, the coercive field is reduced (Fig.2), due to the collective response of the magnetic moments that lead to a reduction of the energy barrier for magnetization reversal. The above features have been also observed in previous studies of FM nanoparticle arrays, where the interface exchange coupling was not included [9, 10]. However, our present calculations indicate that that the reduction of the coercive field with dipolar interactions is sensitive to the strength of the interface exchange. In particular, as shown in Fig.2, a weaker reduction of $H_C$ is observed in systems with stronger interface exchange. This trend is anticipated as a strong J leads to reduced Hc values for the isolated nanoparticles. Nevertheless, the fraction of $H_C$ reduction is nearly the same for both values of J, namely 34% for J=1.0 and 40% for J=1.5.

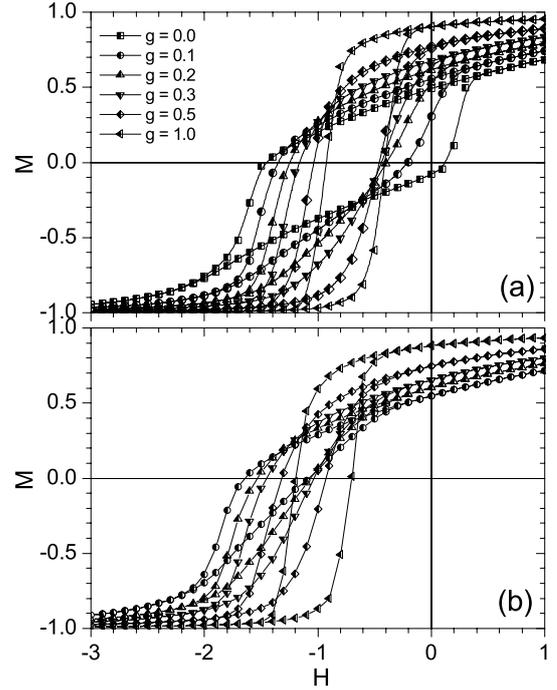

Fig. 1. Low-temperature (T=0.01) isothermal hysteresis of dipolar interacting nanoparticle arrays. Interface exchange interaction (a) J=1.0 and (b) J=1.5. Anisotropy energy $K_C$=1.0 and $K_S$=5.0.

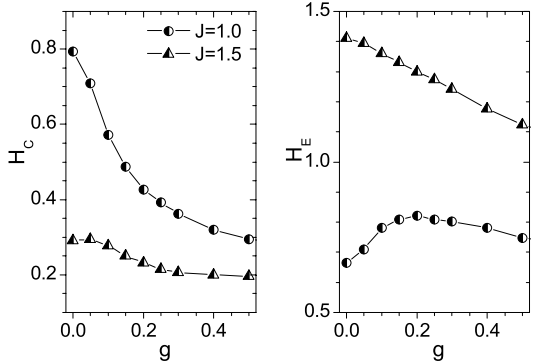

Fig. 2. Dependence of effective coercivity ($H_C$) and exchange bias field ($H_E$) on dipolar coupling, at low temperature (T=0.01). Anisotropy energies $K_C$=1.0 and $K_S$=5.0. Error bars are of the size of the marks.

The behavior of the exchange field with increasing dipolar coupling strength appears more complex (Fig.2). For strong interface coupling (J=1.5) $H_E$ drops linearly with increasing dipolar strength. In the strong dipolar limit (g >>1, not shown here), the exchange field vanishes, as the dominating dipolar forces restore the symmetry of the hysteresis loop around the zero field point. However, a surprisingly different behavior is observed for weaker interface coupling (Fig.2, J=1.0), when the $H_E$ values go through a maximum at intermediate dipolar strength (g≈0.2), before decreasing to zero in the strong dipolar limit. The observed *enhancement* of $H_E$ due to weak dipolar interactions is contrary to what would be expected on intuitive grounds as interactions that lead to symmetric with respect to field-reversal loops, when competing with the random exchange anisotropy field lead to an increased loop shift. We also found a similar increase of $H_E$ with dipolar strength in linear chains of core/shell nanoparticles with random uniaxial anisotropy, while the enhancement disappears in fully textured samples (either linear or planar). We therefore attribute the dipolar-induced enhancement of $H_E$ to the interplay of dipolar interactions and random anisotropy. Finally, we mention that both for interacting and non-interacting arrays, increase of the interface exchange causes reduction of the coercivity and increase of the exchange field, as shown in Fig.2. This behavior is in accordance with atomic scale models [4], where the structural and magnetic details at the interface are taken explicitly into account.

In conclusion, we have shown that in an ordered array of magnetic nanoparticles with a core/shell morphology, the inter-particle dipolar interactions cause suppression of the coercive field, while they produce a more complex behavior of the exchange field. In systems with high shell anisotropy ($K_S$>>$K_C$) and moderate interface coupling (J≈$K_C$), it is found that weak dipolar interactions (g/$K_C$~0.2) could *enhance* the exchange bias effect. This trend is in agreement with recent experiments on stripes of Co/CoO nanoparticles [6], where the strong demagnetizing fields due to the quasi-one-dimensional arrangement of the nanoparticles produced an enhanced exchange field relative to random assemblies. Our simulation results point to the possibility of increasing the exchange field by increasing the packing density. Further work in understanding the details of the magnetization reversal mechanism leading to enhanced $H_E$ values due to dipolar interactions as well as extensions of the MB model to allow for more detailed information concerning the interface structure are currently in progress.

Work supported by EC project NANOSPIN (Contract No NMP4-CT-2004-013545)